# High bandwidth waveguide-integrated plasmonic germanium photodetector

**JACEK GOSCINIAK**[1,*] **AND MAHMOUD RASRAS**[1]

[1]*New York University Abu Dhabi, Saadiyat Island, PO Box 129188, Abu Dhabi, UAE*
*\*jg5648@nyu.edu*

**Abstract:** Here we propose a waveguide-integrated germanium plasmonic photodetector that is based on a long-range dielectric-loaded surface plasmon polariton waveguide configuration. As this configuration ensures a long propagation distance, i.e., small absorption into metal, and a good mode field confinement, i.e., high interaction of the electric field with a germanium material, it is perfect for a realization of plasmonic germanium photodetectors. Such a photodetector even without optimization provides a responsivity exceeding 1 A/W for both wavelengths of 1310 nm and 1550 nm. To achieve such a responsivity, only a 5 μm-long waveguide is required for 1310 nm and 30 μm-long for 1550 nm. With optimization this value can be highly improved. In the proposed arrangement a metal stripe simultaneously supports a propagating mode and serves as one of the electrodes, while the second electrode is located a short distance from the waveguide. As a propagating mode is tightly confined to the germanium ridge, the external electrode can be placed very close to the waveguide without disturbing it. As such, the distance between electrodes can be smaller than 350 nm which allows one to achieve a bandwidth exceeding 100 GHz. However, as most of the carriers are generated inside a distance of 100 nm from a stripe, a bandwidth exceeding 150 GHz can be achieved for a bias voltage of -4 V.

## 1. Introduction

Compared to lasers and modulators that can be combined into one device element, thus avoiding in the optical link, photodetectors are crucial elements that cannot be replaced. They convert light into an electrical signal, making it essential for integrating small but slow electronic components with fast but large sized photonic elements. Furthermore, they are the last elements in the optical link, so they should operate effectively under low input power [1-5]. The optical energy received at a photodetector is directly related to transmitter optical output power and the total link loss. Hence, for 10 fJ/bit transmitted optical energies, the received optical energy would be 1 fJ/bit [3,5]. Therefore, minimizing the optical losses at the photodetector is crucial for the overall performance of the system. Photodetectors usually operate on the basis of the photoelectric effect or exhibit an electrical resistance dependent on the incident radiation. The operation principal is based on the absorption of photons and the subsequent separation of photo-generated charge carriers – electron-hole (e - h) pairs. They suffer, however, from low efficiency either because the photon's energy at telecom wavelengths (0.79 - 0.95 eV) is not sufficient to overcome the Si bandgap (1.12 eV), a low detection area in the case of Ge-based photodetectors (bandgap 0.67 eV), or fabrication problems in the case of graphene-based photodetectors [1-5]. To overcome some of these problems, much attention in recent years has been focused on plasmonics-based photodetectors. Plasmonics photodetectors are attractive because they have the ability to confine light below the diffraction limit, enabling light-matter interaction on a deep sub-wavelength scale [6]. It allows for considerable shrinking of device size, which brings the technology one step closer to the fusion of optical and electronic components of the same size. Plasmonics can serve as a bridge between photonics and electronics by providing components with sizes similar to electronics and speeds characterized by photonics [7,8]. As with all other plasmonic devices, plasmonic photodetectors naturally include metallic elements that can either constitute the absorber in hot-carrier devices, or provide enhancement of the electromagnetic field inside an absorber or both [4,6]. Compared to free space photodetectors, waveguide integrated photodetectors are important components for on-chip optical communication as they can be monolithically integrated with electronics. Furthermore, they can be placed at the terminal end of a waveguide and utilize one of the photo detection schemes.

## 2. State-of-the-art plasmonic photodetectors

In the last few years, the surge of research in waveguide-integrated plasmonic photodetectors has produced results that show very promising performance improvements. In plasmonic photodetectors relying on a hot carrier photodetection schema, a bandwidth of 40 GHz and a responsivity of 0.12 A/W at 1550 nm was measured at a bias voltage of 3.5 V in a metal-insulator-metal (MIM) waveguide arrangement with a footprint below 1 $\mu m^2$ [9]. Another arrangement relies on the inverse-DLSPP waveguide design where a responsivity of 0.085 A/W at 1550 nm wavelength was measured [10,11]. However, by placing a graphene sheet between metal and semiconductor, it is possible to enhance the efficiency of internal photoemission due to a prolonged carrier life-time in the graphene. In this case, a responsivity improvement of 0.37 A/W at 1550 nm wavelength was achieved [12]. Further improvement has been predicted for a thin metal stripe placed entirely inside a semiconductor and operating based on a long-range DLSPP (LR-DLSPP) waveguide arrangement [13]. Taking advantage of this arrangement, a responsivity exceeding 1 A/W was predicted while still keeping bandwidth above 80 GHz. In addition, it has been shown that by replacing noble metals with a TiN as a metal stripe, the improvement in terms of the signal-to-noise ratio (SNR) was predicted as a result of an optimum Schottky barrier height of 0.697 eV. Similar to the inverse-DLSPP, further improvement is possible by placing graphene below or above a metal stripe in LR-DLSPP arrangements.

Additionally, interest has grown in graphene-based photodetectors that can operate based on photo-thermoelectric (PTE), photovoltaic (PC) or bolometric effects [14,15]. Each of these mechanisms may become dominant in different photodetector configurations. Recent work on plasmonic graphene photodetectors utilizing a bolometric effect showed a responsivity of 0.5 A/W at 1550 nm operating at 100 Gbit/s [16]. Another study utilized a narrow asymmetric MIM plasmonic waveguide to provide enhanced light-graphene interaction and enable effectively separate photo-generated carriers [17]. In this case, the responsivity was 0.36 A/W with bandwidth exceeding 110 GHz. In the similar, but symmetric MIM arrangement, another group was able to measure a responsivity of 0.35 A/W and 0.17 A/W for bolometric and photovoltaic effect, respectively [18]. To achieve a responsivity of 0.35 A/W, a bias voltage of only 0.2 V was required [18].

The ideal waveguide-integrated photodetector would need to have high responsivity (quantum efficiency), fast response time, and reduced power consumption, defined by the voltage required to achieve high responsivity. Furthermore, it should be easily implemented into the process flow of photonic foundries. State-of-the-art waveguide-integrated photodetectors need to be at least comparable with available Ge photodetectors that offer a responsivity of 1 A/W at an operating wavelength of 1550 nm. Finally, the bandwidth should exceed 50 GHz and a device should be CMOS compatible. For these reasons, Ge photodetectors are very attractive for potential integration with plasmonic structures.

## 3. Germanium photodetectors

Germanium is very often used in a MSM configuration as it is compatible with the CMOS process and it is a good active material for photodetection in the telecom wavelength range [4,6,19,20]. Germanium belongs to the same group IV materials as Si, so it can be easily integrated with silicon platforms. Compared to Si, which has a relatively large bandgap of 1.12 eV corresponding to an absorption cutoff wavelength of 1100 nm, Ge has a direct bandgap of 0.8 eV, which is only 0.14 eV above the dominant indirect bandgap (0.66 eV), offering a much higher optical absorption in the 1300-1550 nm wavelength range (Fig. 1). This characteristic makes Ge-based photodetectors promising candidates for Si photonic integration [5]. Also, while III-V compound semiconductors possess the advantage of higher absorption efficiency and higher carrier drift velocity, they suffer from integration problems with silicon platforms, increased complexity, and the potential introduction of doping contaminants into Si CMOS devices [5]. Thus, Ge is the more popular material for integration with Si platforms.

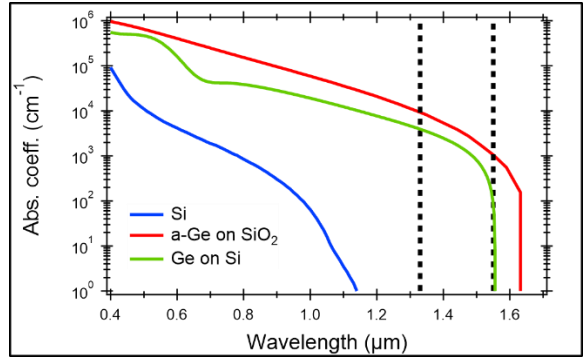

Fig. 1. Absorption coefficients of Si and Ge on Si and SiO2 in wavelengths of 400 – 1700 nm.

Conventional Ge waveguide photodetectors require deposition of metal contacts on Ge. This process, however, introduce a significant losses what influence device responsivity [21-23]. To reduce these losses, new Ge PIN waveguide photodetectors were proposed that exploit lateral Silicon/Germanium/Silicon (Si/Ge/Si) heterojunctions. These results were achieved through improved optical confinement in the Ge layer, yielding a reduction of optical loss in doped contacts. However, the difference in refractive index between doped Si and intrinsic Ge regions is still small, which makes the mode relatively large. As a consequence, for a Ge width of w=0.3 μm, the fraction of optical power inside Ge is estimated to be 54 %, while the fraction of optical power inside doped Si is 32 %. To achieve more power inside Ge, the width of Ge has to be increased. Accordingly, for a Ge width of w=1 μm, around 86 % of optical power stays inside the Ge while only 8% of power is situated in the highly doped Si regions. For a device with Ge width of w=1 μm and a photodetector length of 10 μm, the device responsivity was measured to be 0.5 A/W, which is in good agreement with calculations showing 0.63 A/W. photodetector length of 40 μm, responsivity was measured at 1.2 A/W under low reverse bias voltage of -1 V. It is important to note, however, that for Ge width of 1 μm, the bandwidth hardly reaches 30 GHz under even a high voltage of -3 V, and it is independent of the device length as the frequency response is limited by the carrier transit time. In order to achieve a higher bandwidth, the Ge width needs to be reduced. Combining these results, a 50 GHz bandwidth has been obtained at a reverse bias voltage of 2 V for a Ge width of 0.3 μm and at 3 V for a Ge width of 0.5 μm [22].

To enhance the electric field in the Ge, the metal-semiconductor-metal (MIM) plasmonic waveguide has been proposed where the electric field is highly enhanced in the slot between the two metals [24]. Compared with other SPP waveguide arrangements, the MIM fundamental mode does not exhibit a cut-off, even for very small thicknesses of semiconductor layer. In this way an extremely small mode much below the diffraction limit can be obtained. However, as the gap size reduces, the energy begins to enter the metallic layer, reducing the mode propagation length due to an increase in field localization to the metal-semiconductor interface. As a result, the absorption in the metal arise. Thus, to minimize the losses related with absorption into the metal, it is desired to operate a photodetector in the regime where the absorption in the Ge is high. This way, absorption in the Ge dominates over absorption losses in the metal. The resulting internal quantum efficiency (IQE) is estimated to be 36 % for 1310 nm wavelength and dropping below 10% for the wavelength of 1550 nm. However, to achieve even such results, a very large voltage exceeding 10 V is required. As the MIM field is confined in a narrow Ge region, 100-200 nm, short drift paths for photogenerated carriers are achieved, producing a high speed photoresponse exceeding 100 GHz [24].

## 4. Long-range dielectric-loaded SPP germanium photodetectors

To evaluate the performance of a plasmonic waveguide in terms of applications for a germanium photodetector, the propagation length and mode field confinement should be considered. The former provides information how long the mode can be transmitted, i.e., absorption losses into a metal, and the latter determines the electric field strength inside a waveguide, i.e., interaction between a mode and material.

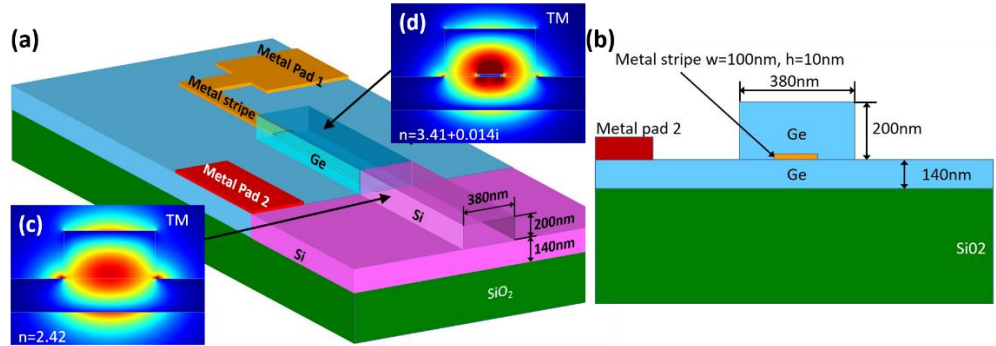

Fig. 2. (a) Proposed Ge LR-DLSPP photodetector arrangement and (b) cross-section of the structure. (c), (d) Mode effective index for (c) Si photonics rib waveguide and (d) Ge LR-DLSPP waveguide.

The operation principle of the proposed photoconductive germanium photodetector is based on the absorption of a long-range dielectric-loaded surface plasmon polariton (LR-DLSPP) mode propagating in a germanium waveguide (Fig. 2). Since the optical energy is perfectly confined inside the waveguide, efficient photodetection process can take place. By applying a bias voltage between the two metallic contacts (metal pad 1 and metal pad 2) (Fig. 2), an electric field is generated in the Ge. Consequently, the generated electron-hole (e-h) pairs are efficiently separated and strongly accelerated by the applied field. These separated carriers drift toward metallic contacts, generating a photo-induced current proportional to the intensity of the optical signal. In the proposed arrangement, the metal stripe supporting a propagating mode is placed between the Ge ridge and Ge slab (Fig. 2b) where the dimensions of ridge and slab were chosen to keep both good mode field confinement and minimum absorption losses in the metal stripe.

### 4.1 Figure of Merit of LR-DLSPP waveguides

LR-DLSPP inherits the advantages of a very long propagation length from long-range SPP (LR-SPP) waveguides and good mode confinement from DLSPP waveguides [25-28]. LR-DLSPP consists of a semiconductor ridge deposited on top of a thin metal stripe/electrode which is supported by another semiconductor slab beneath. The entire structure is supported by a low refractive index substrate which ensures mode confinement to the semiconductor ridge and underlying semiconductor slab. Due to the thin metal film, the SPPs on the two metal-semiconductor interfaces couple to each other and form supermodes with symmetric and anti-symmetric transverse components. The symmetric mode, long-range mode, is characterized by a low longitudinal component of the electric field in the metal, thus lowering absorption losses. The long propagation distance is achieved when effective mode refractive indices on both metal-semiconductor interfaces are close in value to each other so they can couple together to minimize the electric field in the metal. Thus, the LR-DLSPP configuration provides high propagation length and reasonable mode confinement. In comparison, a gap SPP (MIM) can support a very high confined SPP mode but absorption losses from the metal arise, limiting the propagation distance to tens of the corresponding mode's wavelengths. The LR-DLSPP waveguide is currently the only plasmonic waveguide configuration that allows to achieve a good mode field confinement and high propagation length [25-28]. It has the highest evaluated figure of merit (FoM) among all other plasmonic waveguides, taking into account mode size, wavelength, and propagation length (Table 1) [7,26]:

$$FoM = L_p^2 \frac{\lambda_0}{n_{eff} w_0^3}$$

where $w_0$ is the lateral mode width, $L_p$ is the mode propagation length, $n_{eff}$ is the mode effective index, and $\lambda_0$ is the excitation wavelength.

The FoM for LR-DSLPP is at least two orders magnitude higher than other plasmonic waveguide configurations (Table 1) [7,26,29]. Recently, another plasmonic waveguide was proposed, the so-called hybrid photonic-plasmonic waveguide, which offers extremely long propagation length. However, the mode confinement exceeds 7 μm so it is not practical for on-chip integration [29]. Furthermore, the FoM for this waveguide exceeds other plasmonic waveguides, but it still over 30 times lower compared to the LR-DLSPP waveguide. Thus, in terms of the absorption losses into metal, mode field confinement, and possible integration with other on-chip components, the LR-DLSPP waveguide is a perfect candidate for photodetection applications.

Table 1. Figure of merits (FoM) for plasmonic waveguides [7,26].

| Waveguide | LR-DLSPP | LR-SPP | DLSPP | Gap (MIM) | V-grove | HPP [29] |
|---|---|---|---|---|---|---|
| FoM | $3.2 \cdot 10^6$ | $3.2 \cdot 10^4$ | $3.4 \cdot 10^3$ | $1.1 \cdot 10^4$ | $2.9 \cdot 10^4$ | $1.0 \cdot 10^5$ |

The LR-DLSPP mode profile matches very well the mode of the photonic waveguide, allowing for efficient coupling between photonics and plasmonic platforms (Fig. 2c and d). The LR-DLSPP waveguide's mode supporting a TM mode has a similar profile to the photonic TM mode fabricated based on the same material. The overlap integral between them shows up to 98 % coupling efficiency with very good tolerance to the offset of the metal stripe supporting the LR-DLSPP mode [27,28].

The LR-DLSPP configuration therefore demonstrates great potential for creation of photoconductive photodetectors based on germanium and other absorbing materials where good mode confinement and low absorption losses in metal are essential.

### 4.2 LR-DLSPP waveguide for a photodetection

In a state-of-the-art PIN photodetector, the cross-sectional area of the waveguide has to be big enough to minimize absorption losses in a doped Si, so waveguide dimensions of w=1 μm and h=260 nm are required to achieve good mode field confinement in the Ge, low absorption into the doped Si, with high responsivity as a result [21,22]. The cross-sectional area for such a waveguide is 0.26 μm$^2$. In contrast, the plasmonic MIM photodetector offers an extremely small cross-sectional area of 0.016 μm$^2$, but at the cost of absorption losses into the metal that significantly limit its photodetector quantum efficiency [24]. Calculations show that the quantum efficiency of a MIM photodetector operating at 1310 nm does not exceed 70 %, and drops to 30 % for 1550 nm. For the MIM photodetector with Ge between the metals and on top of the MIM structure as presented in Ref. 24, the quantum efficiency does not exceed 10 %.

Compared to the MIM strtucture, the proposed Ge photodetector based on the LR-DLSPP waveguide arrangement provides a cross-sectional area of 0.129 μm$^2$ even when far from optimization. Such a cross-section was chosen to allow efficient coupling of light from the Si waveguide to the photodetector through a butt-coupling arrangement (Fig. 2). In this design, the Ge plasmonic photodetector is an extension of the Si waveguide, enabling an efficient transfer of the optical power to the Ge. Such a design improves device performance in terms of photo-responsivity and opto-electrical bandwidth.

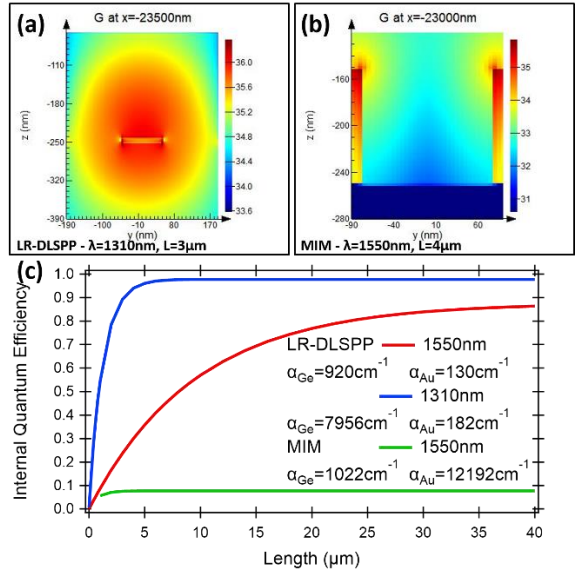

Fig. 3. (a), (b) Generation rate at the middle of the photodetector for (a) LR-DLSPP and (b) MIM photodetector arrangement. (c) Comparison of the losses in metallic stripe/contact on quantum efficiency as a function of the device length for LR-DLSPP and asymmetric MIM arrangements for 1310 nm and 1550 nm wavelengths.

The calculated electric field profile of the fundamental TM mode (Fig. 2d) occupies almost the entire Ge area. Thus, the electron-hole pairs are generated in almost the entire Ge area with most of the electron-hole pairs generated close to the metal stripe that serves simultaneously as one of the electrodes (Fig. 3a). As a result, a quantum efficiency of 97 % can be achieved for a wavelength of 1310 nm even for a very short photodetector length of 5 μm. Furthermore, compared to the MIM plasmonic photodetector that yields quantum efficiencies up to 30 % even for a 40 μm long photodetector operating at 1550 nm, the LR-DLSPP waveguide can achieve a quantum efficiency as high as 84 % for a 30 μm long photodetector operating at the same wavelength of 1550 nm.

## 5. Quantum efficiency of the LR-DLSPP Ge-based photodetector

Neglecting the scattering losses, the responsivity and the internal quantum efficiency (IQE) of a photodetector depends only on the absorption coefficient of Ge, $\alpha_{Ge}$, and metal, $\alpha_m$ [20]:

$$IQE = \frac{\alpha_{Ge} L_{Ge}}{\alpha_{Ge} L_{Ge} + \alpha_m L_m} \left(1 - exp(-(\alpha_{Ge} L_{Ge} + \alpha_m L_m))\right)$$

The materials absorption of Ge and the confinement factor of the mode in the Ge waveguide determines $\alpha_{Ge}$. Also, the overlap of the mode with the metal determines $\alpha_m$ [7]. To calculate the absorption coefficients of metal (Au), $\alpha_{Au}$, and Ge, $\alpha_{Ge}$, 3D FDTD and FEM simulations were performed. To determine $\alpha_{Au}$, the Ge absorption coefficient was set to 0 and the Au stripe/electrode (Fig. 2) was described by a complex refractive index. The reduction in the transmission amplitude through the LR-DLSPP waveguide was assigned to absorption by the Au stripe/electrode. Thus, an effective absorption of Au stripe/electrode, $\alpha_{Au}$, was calculated to be $\alpha_{Au}$=130 cm$^{-1}$ and $\alpha_{Au}$=182 cm$^{-1}$ for wavelengths of λ=1550 nm (Fig. 5) and λ=1310 nm (Fig. 4), respectively. In order to obtain αGe, the absorption coefficient of Au was set to 0, and Ge was described by a complex refractive index, so the losses were assigned to the absorption of the Ge waveguide.

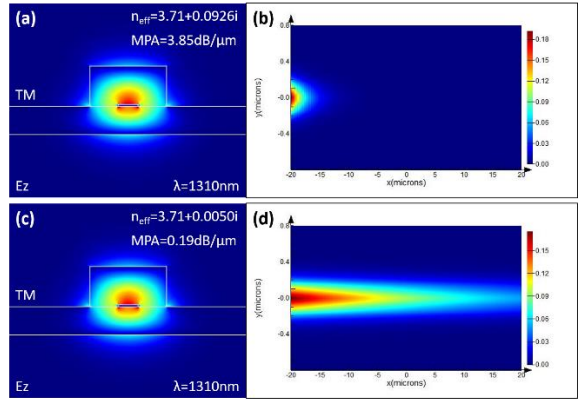

Fig. 4. (a, c) Mode effective index, mode power attenuation (MPA) and (b, d) propagation distance for a structure with complex permittivity of germanium (a, b) and under the assumption of 0 attenuation constant (c, d). Calculation performed for a wavelength of 1310 nm.

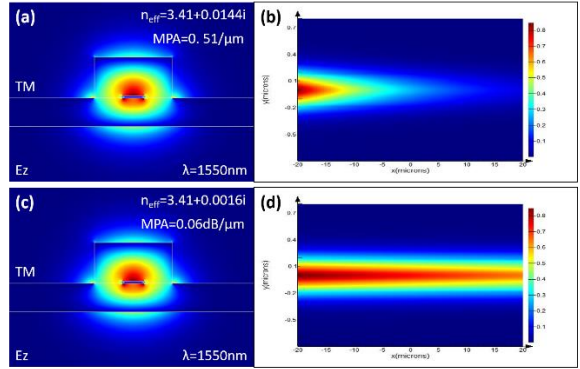

Fig. 5. (a, c) Mode effective index, mode power attenuation (MPA) and (b, d) propagation distance for a structure with complex permittivity of germanium and under the assumption of 0 attenuation constant. Calculation performed for a wavelength of 1550 nm.

The calculated effective absorptions of Ge are $\alpha_{Ge}$=920 cm$^{-1}$ and $\alpha_{Ge}$=7956 cm$^{-1}$ for wavelengths of $\lambda$=1550 nm (Fig. 5) and $\lambda$=1310 nm (Fig. 4), respectively. As shown in Fig. 3c, the quantum efficiencies exceed 97.7 % and 86.3 % for wavelengths of $\lambda$=1310 nm and $\lambda$=1550 nm, respectively. Only a L=24 μm long photodetector is required to achieve a quantum efficiency over 80 % for wavelength $\lambda$=1550 nm, while for wavelength $\lambda$=1310 nm, quantum efficiency of 96 % was possible for the L=5 μm long photodetector. These are record-high results, suggesting the possible realization of a very efficient and fast Ge photodetector. In comparison, a quantum efficiency of 75 % was achieved at $\lambda$=1310 nm for a L=10 μm long Germanium-on-insulator (GOI) MSM photonic photodetector while for $\lambda$=1550 nm it was only 20 % and required a photodetector length of L=20 μm [20]. We can compare our theoretical results with other plasmonic Ge photodetectors realized in the metal-insulator-metal (MIM) plasmonic waveguide arrangement, with Ge placed in a slot of width d=120-160 nm between Au electrodes. In this configuration, a quantum efficiency of 75 % was calculated for the L=10 μm long photodetector at a wavelength of $\lambda$=1310 nm, and only 30 % for the L=50 μm long photodetector at $\lambda$=1550 nm [24]. Compared to these results, our plasmonic photodetector shows a huge improvement in terms of quantum efficiency and footprint. The smaller photodetector's footprint means a lower RC and higher operation speed of the photodetector. The main reason for these improvements can be attributed to good mode field confinement in the entire active area (Fig. 3a) and minimalized losses related with the metal stripe/electrode (Fig. 4, 5).

In the proposed arrangement, the absorption coefficient of Ge, $\alpha_{Ge}$, is at least 7 times higher compared to the absorption coefficient of Au, $\alpha_{Au}$, for a wavelength of $\lambda=1550$ nm, and over 43 times higher for a wavelength of $\lambda=1310$ nm where Ge shows a higher absorption (Fig. 3c). It should be noted, though, that for the other arrangements mentioned previously [20,24], $\alpha_{Ge}$ was always lower than $\alpha_{Au}$. Thus, for GOI photonic waveguides [20], $\alpha_{Ge}$ was 10 times lower compared to $\alpha_{Au}$ for wavelength of $\lambda=1550$ nm and almost the same for wavelength of $\lambda=1310$ nm. Similar trends can be observed for the MIM Ge photodetector [24].

## 6. Responsivity of the LR-DLSPP Ge-based photodetector

The light propagating in the Ge LR-DLSPP waveguide is absorbed mostly by the Ge and partially by a metal (Fig. 3c). The amount of power absorbed by the metal can be almost neglected as it constitutes less than 10 % of overall absorption for a wavelength of 1310 nm and less than 2.5 % for a wavelength of 1550 nm. Here, it is assumed that light is absorbed mostly by the Ge. The absorption per unit volume can be calculated from the divergence of the Poynting vector:

$$P_{abs} = -0.5\text{Re}(\vec{\nabla} \cdot \vec{P}) = -0.5\omega|E|^2\text{Im}(\varepsilon)$$

As a result, an absorption as a function of space and frequency, depends only on the electric field intensity and the imaginary part of the material permittivity. Performed simulations showed that 97 % of the power is absorbed in the first 5 μm of the LR-DLSPP waveguide for a wavelength of 1310 nm and over 88 % of the power coupled to the photodetector is absorbed in the first 30 μm of the photodetector for a wavelength of 1550 nm (Fig. 6a).

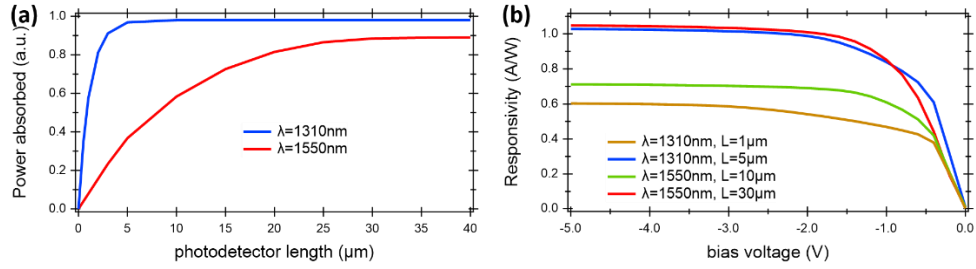

Fig. 6. (a) Total power absorbed by a photodetector (b) Photodetector responsivity as a function of the reserve bias for different device lengths.

These results fit very well with the data in Fig. 3c, where the internal quantum efficiency was calculated analytically based on mode effective index calculations. As can be seen from the above formula, better mode field confinement provides higher electric field intensity, even for the same material, giving rise to higher power absorption. The absorbed photons will generate electron-hole pairs which will then be separated by the electric field applied to the metal electrodes, producing a photocurrent. The number of absorbed photons per unit volume can be calculated by dividing the absorbed power by the Ge by the energy per photon:

$$g = \frac{P_{abs}}{\hbar\omega} = \frac{-0.5\omega|E|^2\text{Im}(\varepsilon)}{\hbar}$$

The frequency-dependent photon absorption rate is equivalent to the generation rate under the assumption that each absorbed photon excites an electron-hole pair. The generation rate for wavelengths of 1310 nm and 1550 nm with different photodetector lengths was calculated using Lumerical FDTD simulations and then exported to Lumerical DEVICES, where it was used to calculate a responsivity for each photodetector configuration (Fig. 6b). As expected from previous calculations (Fig. 3c), the photodetector length of 5 µm is sufficient to produce a responsivity exceeding 1 A/W at 1310 nm. However, the photodetector length needed to obtain a responsivity of 1 A/W at 1550 nm wavelength increases to 30 µm. For a shorter photodetector in the range of 10 µm, the responsivity was calculated to be 0.7 A/W for 1550 nm. These values are higher compared to previously reported results of a plasmonic photodetector based on the asymmetric MIM structure with responsivity measured in the range of 0.05 – 0.1 A/W at the wavelength of 1310 nm [24]. Our calculations confirm this value where the internal quantum efficiency calculated for such a structure was below 0.1 (a.u.) (Fig. 3c). Even assuming a perfect coupling efficiency to the photodetector, the responsivity will not exceed 0.1 A/W [24]. One of the reasons for a much lower responsivity is the absorption losses into the metal that particularly arises for the MIM configuration. Also, as observed in Fig. 3b, the photogenerated electron-hole pairs are generated mostly in the upper side of the MIM structure, with a maximum generation rate localized on both sides of the photodetector in very close proximity to the metal. By comparison, the proposed LR-DLSPP waveguide arrangement generates carriers in the entire area occupied by the germanium waveguide (Fig. 3a). This, together with highly reduced absorption losses caused by metal, allow for responsivity exceeding 1 A/W even for a very compact photodetector.

The dark current achieved in our devices was around 0.6 µA for a 3 µm long photodetector operating at wavelength of 1310 nm. As already shown (Fig. 3c, 6), such a length is sufficient to achieve almost 96 % internal quantum efficiency and responsivity exceeding 1 A/W. This value is much smaller than expected from a typical MSM photodetector. The main reason for this is associated with the small active area of our photodetector. Further improvement can be expected by implementation of asymmetric metallic contacts [34,35] or a thin barrier of a large bandgap material below an external electrode [36].

## 7. Photodetector speed, bandwidth

In telecommunication systems, photodetectors are required to detect optical signals modulated at high data rates. Thus, the important figure of merit is the opto-electrical 3 dB bandwidth, which is defined as the frequency range from DC to cut-off frequency $f_{3dB}$ i.e., the frequency at which the electrical output power drops by 3 dB below power value at very low frequency. The opto-electronic bandwidth of a photodetector depends both on the carrier transit time and RC response time. To reduce the carrier transit time, the distance between the collecting electrodes should be small. In comparison, RC time reduction can be obtained by lowering the contact resistance of the metal electrodes and by reducing the detector length which reduces the photodetector capacitance [30]. The RC-limited 3 dB cut-off frequency can be expressed as:

$$f_{RC} = \frac{1}{2\pi R_{eff} C_{pd}}$$

where $R_{eff}$ is the effective overall resistance and $C_{pd}$ is the junction capacitance. The carrier transit time, that defines the time of photo-generated electrons or hole to travel through the active region prior to being collected by the contacts, can be estimated using:

$$f_t \approx \frac{3.5v}{2\pi d_{abs}}$$

where $d_{abs}$ is the distance between the site where carriers are generated and the electrode collecting those carriers, and v is the average carrier velocity [13,30,37],

As mentioned above, only a L=5 µm long photodetector is required to achieve a quantum efficiency exceeding 95 % at a wavelength of 1310 nm. Taking into account the distance between both electrodes to be approximately 450 nm i.e., the second electrode is placed around 300 nm from the Ge waveguide, a capacitance in the range of a few fF is suspected. The second electrode placed 300 nm from the waveguide would not disturb a propagating mode in the waveguide (Fig. 8c). As a result, there will be small propagation losses and associated with it small absorption by the metal electrode supporting the propagating mode, giving rise to enhanced quantum efficiency and responsivity. If the device/photodetector is connected with a 50 Ω load, a RC cut-off frequency exceeding a 1 THz can be achieved.

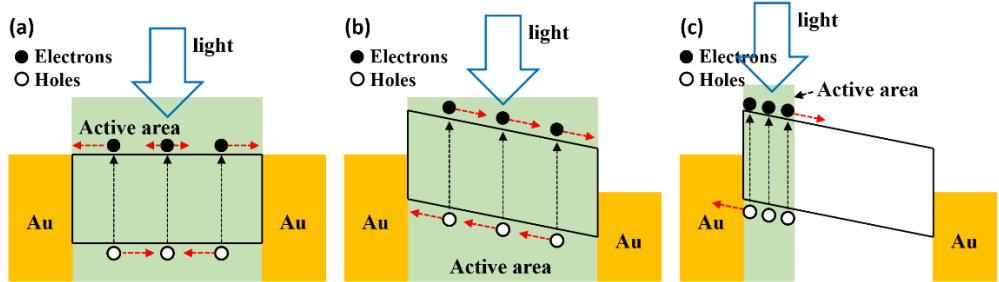

Fig. 7. Schematic of the band diagram of the Au-Ge-Au structure (a) without bias and (b, c) under bias for a structures (a, b) with the active area extended between both metal contacts and (c) limited only to the small volume around the electrode supporting a propagating LR-DLSPP mode.

In the proposed design, all carriers are generated in the area limited by the germanium waveguide. Assuming a distance between electrodes of 450 nm and a drift velocity of $6.5 \cdot 10^6$ cm/s [37], a bandwidth of $f_t=80$ GHz can be achieved. However, as it can be observed from Fig. 4 and 5, over 90 % of absorbed power is located around 100 nm from the collecting metal electrode.

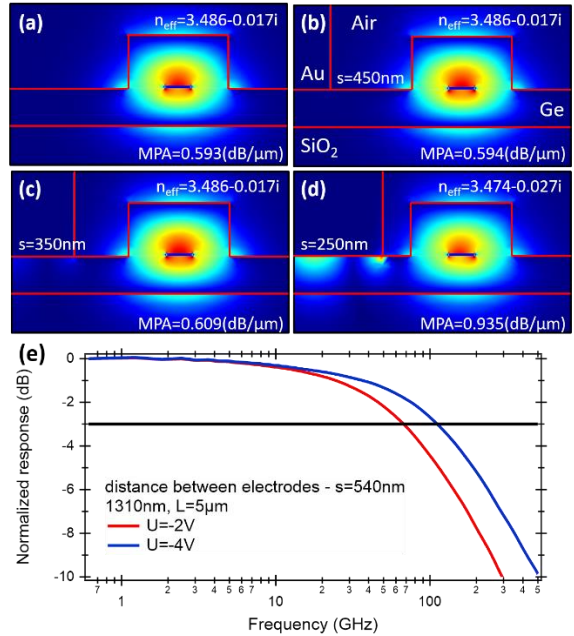

Fig. 8. (a)-(d) Mode effective index and corresponding MPAs for different spacing between the electrodes. (e) Normalized frequency responses under -2 V and -4 V reverse bias voltage for electrode spacing of s=540 nm.

Therefore, a bandwidth exceeding 100 GHz can be achieved even for an electrode spacing of 450 nm, which shows potential for the realization of high bandwidth photodetectors based on the LR-DLSPP waveguide arrangement (Fig. 8e). Because of the nature of SPPs that enables strong light concentration in the near-field, most of the carriers are generated in very close proximity to the metal stripe/electrode. As a result, the average photogenerated electron path length to the metal electrode is significantly reduced in comparison to the conventional photoconductive photonic photodetectors [31]. Hence, the photogenerated carriers reach a metal electrode much faster even in the case of absorbant materials with relatively low drift velocity in the semiconductor. Furthermore, a big proportion of the photogenerated electrons can reach the contact in a sub-picosecond timescale, which improves the operation bandwidth and the photodetector's efficiency and responsivity. It should be noted that in conventional photoconductive photodetectors the majority of the photo-generated carriers recombine in the substrate before reaching the contact electrodes [31]. With the characteristics described above, the proposed photodetector can offer a bandwidth exceeding 150 GHz.

As previously shown, the bandwidth of the photodetector is not limited by the RC time constant but the transit time required for carriers to drift to the contact electrode collecting the carriers. By decreasing the distance between the two contacts, a very high electric field is obtained at low bias voltage what additionally improves carrier collection.

Our proposed Ge plasmonic photodetector offers a unique combination of advantages and a huge flexibility in terms of design and integration with other materials. Due to the difference in velocity of the carriers, i.e., electrons and holes, the response of the photodetector is usually asymmetric as holes are delayed compared to electrons. However, in our proposed LR-DLSPP photodetector arrangement, carriers are photogenerated in the region surrounding the metal stripe that serves simultaneously as a collection electrode (Fig. 2 and Fig 7c). Consequently, the holes with lower mobility and shorter paths are only moving to the metal stripe were they are collected whereas the electrons move to the external electrode. Furthermore, this design allows for further improvement through combining it with narrow and wide bandgap semiconductors such as n-Si, so electrons can be isolated and the photoresponse of the device will only be dependent on the transport of electrons.

## 8. Symmetric and asymmetric electrodes

Compared to previous waveguide arrangements where carriers were generated in the entire area limited only by metal contacts (Fig. 7a, b) [20-24] (without a bias voltage (Fig. 7a) and under a bias voltage (Fig. 7b)), in the proposed arrangement the carriers are only generated close to the metal electrode/contact supporting a propagating mode (Fig. 7c, under a bias voltage). This results in an extremely fast transit time for carriers to reach the electrode. The metal electrodes/contacts can consist of the same metals creating a symmetric structure, or they can create an asymmetric structure when two different metals are used. In terms of the metal electrodes supporting a propagating mode, the electrode should show good plasmonic properties defined by the real and imaginary parts of the permittivity. The real part of permittivity should be negative with as large a value as possible. Simultaneously, to minimize propagation losses related to absorption, the imaginary part of permittivity should be as small as possible in the desired wavelength range. In the last few years, a new class of plasmonic materials appeared, so-called transition metal nitrides (TMNs), which offer optical properties close to those of noble metals [13,32,33]. Additionally, they show high temperature durability and CMOS compatibility while their optical properties strongly depend on the deposition conditions [13,32,33]. As TMNs, such as for example TiN, are CMOS compatible, they can be integrated in the proposed LR-DLSPP photodetector waveguide arrangement giving more flexibility in the design process. This would allow one to design a photodetector with asymmetric metallic contacts, where TiN can be arranged either as an external electrode (metal pad 2, Fig. 2) or a metal electrode supporting a propagating plasmonic mode and connected with metal pad 1 (Fig. 2). Furthermore, other materials with good plasmonic properties, such as aluminum, copper, silver, gold, zirconium nitride (ZrN) and others, can be used as a metal stripes supporting a propagating mode. In the case of an external electrode, all materials showing good electrical properties can be implemented in this arrangement. Previous work has shown that asymmetric metallic contacts are able to minimize a dark current in the photodetector [34,35]. An alternative method is by adding a thin barrier with a large bandgap material between metal contacts and germanium [36].

## 9. Summary

In this study we have proposed the first plasmonic germanium photodetector operating at telecom wavelengths of 1310 nm and 1550 nm with a record high responsivity exceeding 1 A/W and bandwidth beyond 100 GHz for both wavelengths. An internal quantum efficiency exceeding 98 % was achieved for only a 5 μm long photodetector at a wavelength of 1310 nm, while for a wavelength of 1550 nm the quantum efficiency was calculated to be 84 % for 30 μm long photodetector. Both of these values are much higher than previously reported for both plasmonic and photonic germanium photodetectors. These results have been achieved owing to a long-range dielectric-loaded SPP waveguide configuration that ensures good mode field confinement and low absorption losses into a metal. Finally, the proposed photodetector arrangement can be easily integrated with silicon photonic platforms enabling on-chip integration with other components.

*Acknowledgments*

Support from the NYUAD Research Grant is gratefully acknowledged.

*Disclosure*

The authors declare that there are no conflicts of interest related to this article.

*Contribution*

J.G. conceived the idea of the paper, devised the theoretical model and performed the computer simulations and wrote the manuscript. J.G. and M.R. reviewed the manuscript.

## 10. References

1. D. A. B. Miller, "Attojoule Optoelectronics for Low-Energy Information Processing and Communications," J. of Lightw. Technol. 35(3), 346-396 (2017).
2. L. C. Kimerling, D-L Kwong, and K. Wada, "Scaling computation with silicon photonics," MRS Bulletin 39, 687-695 (2014).
3. D. Thomson at el., "Roadmap on silicon photonics," J. of Optics 18, 073003 (2016).
4. Ch. A. Thraskias, at al., "Survey of Photonic and Plasmonic Interconnect Technologies for Intra-Datacenter and High-Performance Computing Communications," IEEE Comm. Surveys & Tutorials 20(4), Fourth quarter (2018).
5. M. Piels and J. E. Bowers, "1 – Photodetectors for silicon photonic integrated circuits," Editor(s): B. Nabet, "Photodetectors," Woodhead Publishing, 3-20 (2018).
6. A, Dorodnyy et al., "Plasmonic Photodetectors," IEEE J. of Selected Topics in Quantum Electronics 24(6), 4600313 (2018).
7. Z. Han, and S. I. Bozhevolnyi, "Radiation guiding with surface plasmon polariton," Rep. Prog. Phys. 76, 016402 (2013).
8. A. Kumar, et al., "Dielectric-loaded plasmonic waveguide components: going practical," Laser & Photonics Reviews 7 (6), 938-951 (2013).
9. S. Muehlbrandt, et al., "Silicon-plasmonic internal-photoemission detector for 40 Gbit/s data reception," Optica 3 (7), 741-747 (2016).
10. I. Goykhman, B. Desiatov, J. Khurgin, J. Shappir, and U. Levy, "Locally oxidized silicon surface-plasmon Schottky detector for telecom regime," Nano Lett. 11(6), 2219-2224 (2011).
11. I. Goykhman, B. Desiatov, J. Khurgin, J. Shappir, and U. Levy, "Waveguide based compact silicon Schottky photodetector with enhanced responsivity in the telecom spectral band," Opt. Express 20(27), 28594-28602 (2012).
12. I. Goykhman, et al., "On-chip integrated, silicon–graphene plasmonic Schottky photodetector with high responsivity and avalanche photogain," Nano Lett. 16(5), 3005-3013 (2016).
13. J. Gosciniak, F. B. Atar, B. Corbett, and M. Rasras, "Plasmonic Schottky photodetector with metal stripe embedded into semiconductor and with a CMOS-compatible titanium nitride," Sci. Rep. 9, 6048 (2019).
14. T. J. Echtermeyer, et al., "Photo-thermoelectric and photoelectric contributions to light detection in metal-graphene-metal photodetectors," Nano Lett. 14, 3733–3742 (2014).
15. M. Freitag, T. Low, F. Xia, and P. Avouris, "Photoconductivity of biased graphene," Nat. Photon. 7, 53–59 (2012).

16. P. Ma, Y. Salamin, B. Baeuerle, A. Josten, W. Heni, A. Emboras, and J. Leuthold, "Plasmonically Enhanced Graphene Photodetector Featuring 100 Gbit/s Data Reception, High Responsivity, and Compact Size," ACS Photonics 6(1), 154-161 (2018).
17. Y. Ding, et al., "Ultra-compact integrated graphene plasmonic photodetector with bandwidth above 110 GHz", arXiv:1808.04815 (2018).
18. Z. Ma, et al., "Compact Graphene Plasmonic Slot Photodetector on Silicon-on-insulator with High Responsivity", arXiv:1812.00894 (2018).
19. G. Li et al., "Improving CMOS-compatible Germanium photodetectors," Opt. Express 20(24), 26345-26350 (2012).
20. S. Assefa, et al., "CMOS-integrated high-speed MSM germanium waveguide photodetector," Opt. Express 18(5), 4986-4999 (2010).
21. L. Virot, et al., "Integrated waveguide PIN photodiodes exploiting lateral Si/Ge/Si heterojunction," Opt. Express 25(16), 19487 (2017).
22. D. Benedikovic, et al., "25 Gbps low-voltage hetero-structured silicon-germanium waveguide pin photodetectors for monolithic on-chip nanophotonic architectures," Photonics Research 7(4), 437 (2019).
23. H. Chen, et al., "100-Gbps RZ Data Reception in 67-GHz Si-Contacted Germanium Waveguide p-i-n Photodetectors," J. of Lightw. Technol. 35(4) (2017).
24. Y. Salamin, et al., "100 GHz Plasmonic Photodetector," ACS Photonics 5(8), 3291-3297 (2018).
25. T. Holmgaard, J. Gosciniak, and S. I. Bozhevolnyi, "Long-range dielectric-loaded surface plasmon-polariton waveguides," Opt. Express 18(22), 23009-23015 (2010).
26. J. Gosciniak, T. Holmgaard, and S. I. Bozhevolnyi, "Theoretical Analysis of Long-Range Dielectric-Loaded Surface Plasmon Polariton Waveguides," J. of Lightw. Technol. 29(10), 1473-1481 (2011).
27. B. Sturlesi, M. Grajower, N. Mazurski, and U. Levy, "Integrated amorphous silicon-aluminium long-range surface plasmon polariton (LR-SPP) waveguides," APL Photonics 3, 036103 (2018).
28. X. Shi, X. Zhang, Z. Han, U. Levy, and S. I. Bozhevolnyi, "CMOS-Compatible Long-Range Dielectric-Loaded Plasmonic Waveguides," J. of Lightw. Technol. 31(21), 3361-3367 (2013).
29. S. Saha, et al., "On-Chip Hybrid Photonic-Plasmonic Waveguides with Ultrathin Titanium Nitride Films," ACS Photonics 5, 4423−4431 (2018).
30. H. Venghaus and N. Grote, "Fibre Optic Communication: Key Devices," 2nd Edition, Springer Series in Optical Sciences (2017).
31. C. W. Berry "Significant performance enhancement in photoconductive terahertz optoelectronics by incorporating plasmonic contact electrodes," Nat. Comm. 4, 1622 (2013).
32. P. R. West, S. Ishii, G. V. Naik, N. K. Emani, V. M. Shalaev, and A. Boltasseva, "Searching for better plasmonic materials," Laser & Photonics Reviews 4(6), 795–808 (2010).
33. J. Gosciniak. J. Justice, U. Khan, M. Modreanu, and B. Corbett, "Study of high order plasmonic modes on ceramic nanodisks," Opt. Express 25(5), 5244-5244 (2017).
34. C. O. Chui, A. K. Okyay, and K. C. Saraswat, "Effective dark current suppression with asymmetric MSM photodetectors in Group IV semiconductors," IEEE Photon. Technol. Lett. 15(11), 1585–1587 (2003).
35. H. –J. Zang, et al., "Asymmetrically contacted germanium photodiode using a metal – interlayer – semiconductor − metal structure for extremely large dark current suppression," Opt. Letters 41 (16), 3686−3689 (2016).
36. J. –Y. Lin, A. M. Roy, A. Nainani, Y. Sun, K. C. Saraswat, "Increase in current density for metal contacts to n-germanium by inserting $TiO_2$ interfacial layer to reduce Schottky barrier height," Appl. Phys. Lett. 98 (9), 092113 (2011).
37. Yi Zhang, et al., "A high-responsivity photodetector absent metal germanium direct contact," Opt. Express 22(9), 011367 (2014).